%% file: revision1.tex
\documentclass[12pt,twoside]{article}   
\usepackage[super,sort,comma]{natbib}

\usepackage{ulem}
\usepackage{fancyhdr}		

\usepackage[section]{placeins}   %

\usepackage[utf8]{inputenc} 
\usepackage[T1]{fontenc}    
\usepackage{hyperref}       
\usepackage{url}            
\usepackage{booktabs}       
\usepackage{amsfonts}       
\usepackage{nicefrac}       
\usepackage{microtype}      
\usepackage[table]{xcolor}         
\usepackage{pifont}

\usepackage{marvosym}
\usepackage{amsmath}
\usepackage{amssymb}
\usepackage{multirow}
\usepackage{arydshln}
\usepackage{graphicx}
\usepackage{subcaption}

\makeatletter \renewcommand\@biblabel[1]{$^{#1}$} \makeatother
\newcommand{\cen}[1]{\begin{center} #1 \end{center}}

\usepackage[mathlines]{lineno}

\usepackage{hyperref}
\hypersetup{ colorlinks,
    citecolor=blue,
    filecolor=blue,
    linkcolor=blue,
    urlcolor=blue
}

\usepackage{xcolor}

\definecolor{gray}{rgb}{0.6,0.6,0.6}
\definecolor{red}{rgb}{0.85,0,0}
\definecolor{green}{rgb}{0,0.85,0}
\definecolor{blue}{rgb}{0,0,0.85}
\definecolor{beige}{rgb}{0.92,0.87,0.78}
\usepackage[all]{hypcap}    

\begin{document}
\cen{\sf {\Large {\bfseries Continuous sPatial-Temporal Deformable Image Registration and 4D Frame Interpolation} \\  
\vspace*{15mm}
Xia Li\textsuperscript{1,2}, Runzhao Yang\textsuperscript{2,4}, Muheng Li\textsuperscript{1,3}, Xiangtai Li\textsuperscript{5}, \hspace{2cm}Antony J. Lomax\textsuperscript{1,3}, Joachim M. Buhmann\textsuperscript{2}, Ye Zhang\textsuperscript{1}}\\
\vspace{5mm}
1. Center for Proton Therapy, Paul Scherrer Institut, 5232 Villigen PSI, Switzerland \\
2. Department of Computer Science, ETH Zürich, 8092 Zürich, Switzerland \\
3. Department of Physics, ETH Zürich, 8092 Zürich, Switzerland \\
4. Department of Automation, Tsinghua Univeristy, China \\
5. S-Lab, Nanyang Technological University, Singapore \\
\vspace{5mm}
Version typeset \today\\
\vspace{5mm}
}

\pagenumbering{roman}
\setcounter{page}{1}
\pagestyle{plain}
\sf Author to whom correspondence should be addressed. Email: ye.zhang@psi.ch \\

\begin{abstract}
\noindent{\bf Background:} 
Deformable image registration (DIR) is a crucial tool in radiotherapy for analyzing anatomical changes and motion patterns. Current DIR implementations rely on discrete volumetric motion representation, which often leads to compromised accuracy and uncertainty when handling significant anatomical changes and sliding boundaries. This limitation affects the reliability of subsequent contour propagation and dose accumulation procedures, particularly in regions with complex anatomical interfaces such as the lung-chest wall boundary.

\noindent{\bf Purpose:}
Given that organ motion is inherently a continuous process in both space and time, we aimed to develop a model that preserves these fundamental properties. Drawing inspiration from fluid mechanics, we propose a novel approach using implicit neural representation (INR) for continuous modeling of patient anatomical motion. This approach ensures spatial and temporal continuity while effectively unifying Eulerian and Lagrangian specifications to enable natural continuous motion modeling and frame interpolation. The integration of these specifications provides a more comprehensive understanding of anatomical deformation patterns.

\noindent{\bf Methods:} 
We propose an implicit neural representation-based approach modeling motion continuously in both space and time, named Continues-sPatial-Temporal Deformable Image Registration (CPT-DIR). This method fits a multilayer perception network to map the 3D coordinate $(x, y, z)$ to its corresponding velocity vector $(vx, vy, vz)$. Displacement vectors $(\triangle x, \triangle y, \triangle z)$ are then calculated by integrating velocity vectors over time using an Euler method numerical scheme. The above spatial and temporal continuous motion design also enables continuous frame interpolation (CPT-Interp). The DIR's and interpolation's performance were tested on the DIRLab dataset, using metrics of landmark accuracy (TRE), contour conformity (Dice), and image similarity (MAE).

\noindent{\bf Results}: 
CPT-DIR significantly reduced landmark TRE from $2.79 \pm 1.88$ mm to $0.99 \pm 1.07$ mm, surpassing B-spline results across all cases. The whole-body region MAE improved from $35.46 \pm 46.99$ HU to $28.99 \pm 32.70$ HU. In the challenging sliding boundary region, CPT-DIR demonstrated superior performance compared to B-spline, reducing ribcage MAE from $75.40 \pm 86.70$ HU (unregistered) to $42.04 \pm 45.60$ HU and improving Dice coefficients from $89.30\%$ to $90.56\%$. The training-free CPT-Interp method enhanced previous deep learning-based approaches, improving upon UVI-Net with reduced MAE ($17.88 \pm 3.79$ versus $18.93 \pm 3.90$) and increased PSNR ($40.26 \pm 1.58$ versus $39.76 \pm 1.48$), while eliminating training dataset dependencies. Both CPT-DIR and CPT-Interp achieved substantial computational efficiency, completing operations in under 3 seconds compared to several minutes required by conventional B-spline methods.

\noindent{\bf Conclusion:} 
By leveraging the continuous representations, the CPT-DIR method significantly enhances registration and interpolation accuracy, automation, and speed. The method demonstrates superior performance in landmark and contour precision, particularly in challenging anatomical regions, representing a substantial advancement over conventional approaches in deformable image registration. The improved efficiency and accuracy of CPT-DIR make it particularly suitable for real-time adaptive radiotherapy applications.

\end{abstract}

\newpage     

\tableofcontents

\newpage

\setlength{\baselineskip}{0.7cm}      

\pagenumbering{arabic}
\setcounter{page}{1}
\pagestyle{fancy}
\input{sections-r1/1.intro}
\input{sections-r1/2.methods}
\input{sections-r1/3.results}

\input{sections-r1/4.discuss}
\input{sections-r1/5.conclusion}
\section*{Acknowledgments}
\addcontentsline{toc}{section}{\numberline{}Acknowledgments}
This work was supported by the Personalized Health and Related Technologies (PHRT) of the ETH Domain, Switzerland through an interdisciplinary doctoral grant (iDoc 2021-360), and by the Swiss National Science Foundation (SNSF) under Grant No. 212855.

\section*{Appendix}
\addcontentsline{toc}{section}{\numberline{}Appendix}
\input{sections-r1/6.appendix}

\clearpage
\section*{References}
\addcontentsline{toc}{section}{\numberline{}References}
\vspace*{-10mm}





\bibliography{./example}      



\bibliographystyle{./medphy.bst}    


\end{document}

%% file: sections-r1/1.intro.tex
\section{Introduction}
In radiotherapy, accurate alignment of anatomical structures is of paramount importance for ensuring high precision treatment delivery and minimizing damage to healthy tissues~\cite{verhey1995immobilizing}. 
Deformable image registration (DIR) is a key technique used to achieve this alignment by estimating the dense displacement vector field (DVF) between sequential images, thereby enabling the mapping of anatomical structures from one image to another~\cite{oh2017deformable,nenoff2023review}. 
DIR plays a crucial role in various aspects of radiotherapy, including contour propagation, dose accumulation, and 4D dose calculation ~\cite{rigaud2019deformable,zhang2012respiratory}.
However, achieving robust and precise DIR in highly dynamically changed anatomical regions, such as the head-and-neck, remains a significant challenge~\cite{mok2020large}. These regions are subject to complex inter-fraction and intra-fraction anatomical changes due to variations in patient positioning, respiration, and tissue deformation~\cite{amstutz2024quantification}. Such changes can lead to substantial discrepancies between the planning and treatment images, compromising the accuracy of dose delivery and potentially leading to suboptimal treatment outcomes. Previous DIR methods, which rely on discrete spatial modeling with end-to-end mapping, often struggle to handle these complex anatomical variations.

The complexity of DIR arises from several factors, including the lack of definitive ground truth~\cite{fischer2008ill} and the high degree of freedom in DVF optimization~\cite{sotiras2013deformable}. Over the years, DIR methods have evolved from traditional approaches, such as optical flow~\cite{ostergaard2008acceleration,yang2008fast} and elastic models~\cite{kybic2003fast,davatzikos1997spatial}, to more advanced techniques, such as the Demons algorithm~\cite{vercauteren2009diffeomorphic,vercauteren2007non} and B-spline registration~\cite{rueckert2006diffeomorphic,klein2007evaluation}. These advancements have improved the performance of DIR in terms of accuracy and efficiency; however, they still have limitations, especially the slow speed~\cite{de2019deep} and expert tuning~\cite{fu2020deep}.
More recently, deep learning-based approaches have emerged as a promising direction for advancing DIR~\cite{haskins2020deep}. Methods like VoxelMorph~\cite{balakrishnan2019voxelmorph} have demonstrated the potential of unsupervised learning in improving the accuracy and speed of DIR by learning the DVF directly from image data. SynthMorph~\cite{hoffmann2021synthmorph} has extended these capabilities to multi-modal image registration, enabling the alignment of images from different modalities, such as CT and MRI. Transmorph~\cite{chen2022transmorph} has further enhanced spatial correspondence modeling by incorporating transformer models, which can capture long-range dependencies and improve the robustness of DIR. Despite these advancements, deep learning-based DIR methods still face challenges, such as the requirement for large and diverse training datasets and the difficulty in generalizing to new data or imaging modalities~\cite{zajkac2023ground}.

Beyond the DIR task, several methods have been proposed, specifically 4D medical image interpolation~\cite{guo2021unsupervised}. SVIN~\cite{guo2020spatiotemporal} employed a dual-network strategy for capturing and interpolating volumetric motion, but its application was limited by its reliance on extensive radiation, lengthy imaging processes, and the availability of ground-truth intermediate images for training. MPVF~\cite{wei2023mpvf} used multi-pyramid voxel flows to address periodic motion in organ structures; however, it struggled with non-periodic motions due to its discrete representation of dynamic biological processes. UVI-Net~\cite{kim2024data} avoids using intermediate frames through a cycle consistency model, enhancing image fidelity from limited data. Yet, its reliance on linear motion assumptions can lead to spatial distortion and inaccuracies in capturing complex physiological movements. Despite these advancements, accurately modeling anatomic changes' continuous and complex nature over time remains an ongoing challenge.

To overcome these limitations, we propose an advanced deformation modeling approach that leverages continuous spatial and temporal representations for precise registration. Our method employs Implicit Neural Representations (INR)~\cite{molaei2023implicit,sitzmann2020implicit,wolterink2022implicit} to create continuous models of DVFs that capture the fine-grained details of anatomical deformations over time and space. By utilizing INR, we aim to move beyond conventional voxel-based registration techniques, offering a more flexible and accurate way to model complex deformations during proton therapy. We model the registration process as a spatial-temporal continuous flow using a Multi-Layer Perceptron (MLP) network, effectively addressing sliding boundary issues and large deformations while adapting quickly to new cases without extensive hyper-parameter tuning.
The main contributions of this work are threefold. First, we introduce a novel INR-based approach for modeling spatial-temporal deformations in medical image registration. Second, we demonstrate the effectiveness of our method in capturing complex anatomical changes in highly dynamic regions, such as the lung. Finally, we compare our approach with state-of-the-art DIR and 4D medical image interpolation methods, highlighting its advantages in terms of accuracy and robustness.

\noindent

%% file: sections-r1/2.methods.tex
\section{Methods}\label{sec:4.1}
\subsection{Conventional Approaches for DIR and Interpolation}
\definecolor{myblue}{RGB}{0,112,192}
\definecolor{myred}{RGB}{192,0,0}
\begin{figure}[ht]
    \centering
    \begin{minipage}[c]{0.32\textwidth}
        \centering
        \includegraphics[width=\textwidth]{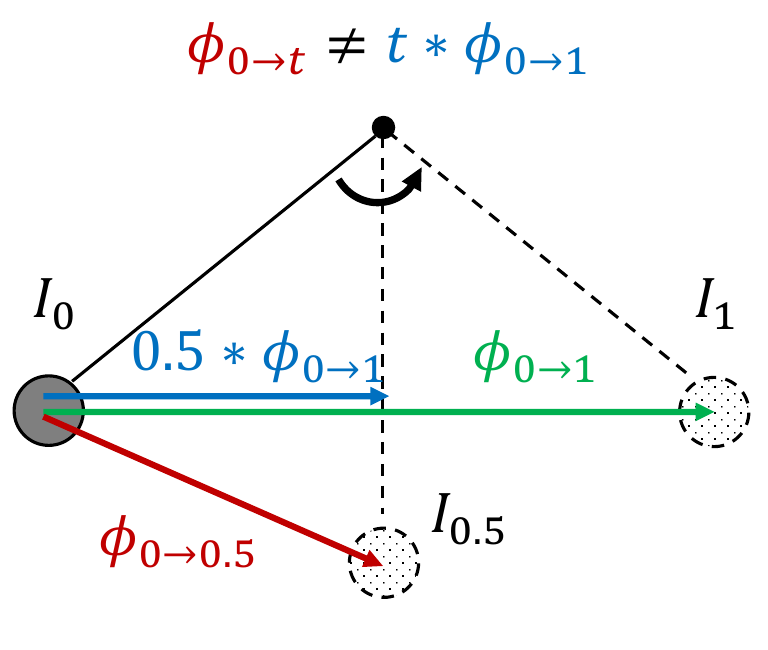}
        \subcaption{Nonlinear motion.}
        \label{fig:4.1a}
    \end{minipage}
    \begin{minipage}[c]{0.32\textwidth}
        \centering
        \includegraphics[width=\textwidth]{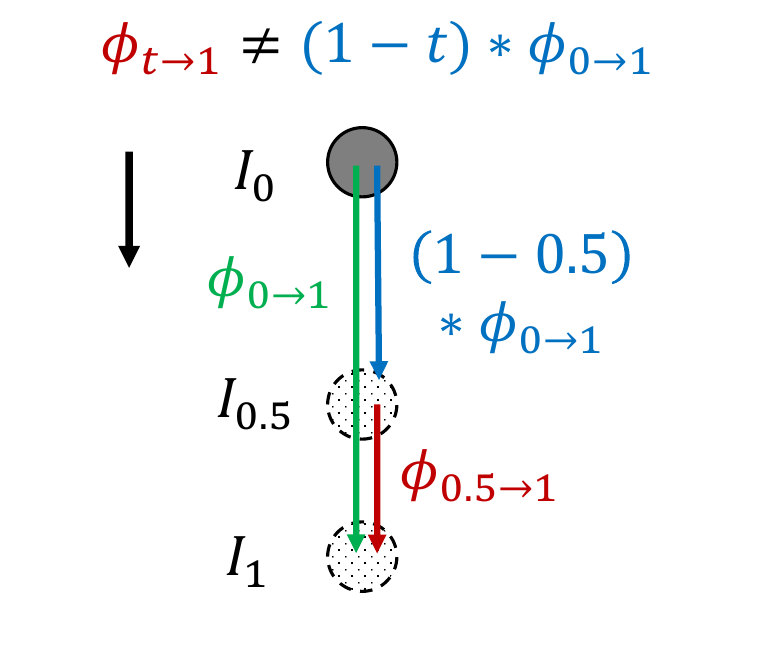}
        \subcaption{Linear motion.}
        \label{fig:4.1b}
    \end{minipage}
    \begin{minipage}[c]{0.32\textwidth}
        \centering
        \includegraphics[width=\textwidth]{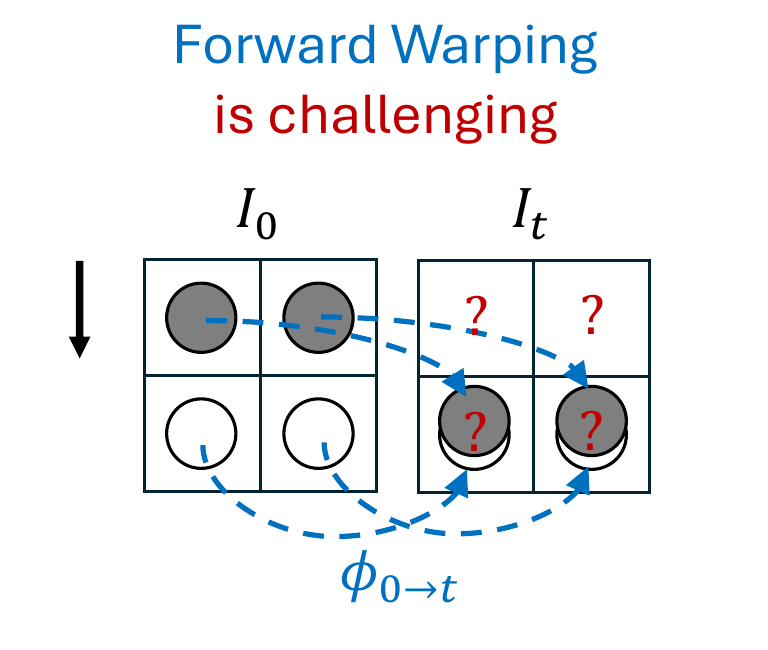}
        \subcaption{Forward warping.}
        \label{fig:4.1c}
    \end{minipage}
    \caption{(a) In the nonlinear motion shown, \textcolor{myblue}{$t*\phi_{0\to1}$} does not give an accurate \textcolor{myred}{$\phi_{0\to t}$}. They start at the same point, but there are errors in direction and magnitude. (b) In the linear motion shown, \textcolor{myblue}{$(1-t)*\phi_{0\to1}$} does not give an accurate \textcolor{myred}{$\phi_{t\to 1}$} either. They are in the same direction, but the modeling errors are in the starting point and magnitude. (c) In the overlapping scenario shown, even with an accurate \textcolor{myblue}{$\phi_{0\to t}$}, ambiguities arise during the forward warping of $I_0$ : i) What values should be assigned to the \textcolor{myred}{two upper pixels} that are not targeted by the warping? ii) What values should be assigned to the \textcolor{myred}{two lower pixels} that are mapped from multiple sources?} 
    \label{fig:4.1}
    \vspace{5mm}
\end{figure}
Both deformable image registration and 4D frame interpolation rely on motion modeling, typically achieved by estimating the displacement field $\phi_{0 \to 1}$, which is linear in time.
Given a pair of 3D images $I_0$ (fixed) and $I_1$ (moving), where voxel $z_0 \in \mathbb{Z}^3$, a DIR algorithm estimates the DVF $\phi_{0 \to 1}$ that maps the coordinates in $I_0$ to their corresponding locations in $I_1$, aligning the two images by minimizing the difference between the reference image $I_0$ and the warped moving image $I_1 \circ \phi_{0 \to 1}$. Since this problem is ill-posed in general, regularization terms are introduced to constrain the solution space of $\phi_{0 \to 1}$. The optimization problem for DIR can be formulated as:
\begin{equation}
    \phi^*_{0 \to 1} = \underset{\phi_{0 \to 1}}{\arg\min} D(I_0, I_1 \circ \phi_{0 \to 1}) + \lambda R(\phi_{0 \to 1}),
\end{equation}
where $D$ measures the image similarity and is usually implemented as the Sum of Squared Differences (SSD)~\cite{hisham2015template}, Normalized Cross Correlation (NCC)~\cite{zhao2006image}, or Normalized Gradient Fields (NGF)~\cite{konig2014fast}. Additionally, $R$ with strength $\lambda$ is the regularization term; possible choices for $R$ include the classic L1 loss, Total Variation (TV) loss~\cite{vishnevskiy2016isotropic}, etc. Conventional methods directly optimize the DVF map $\phi_{0 \to 1}$ or use condensed representations like B-spline~\cite{zhang1994iterative}, while deep learning-based methods~\cite{balakrishnan2019voxelmorph,hoffmann2021synthmorph,chen2022transmorph,rueckert2006diffeomorphic,vercauteren2009diffeomorphic} learn a mapping function $f: I_0 \times I_1 \to \phi_{0 \to 1}$ from a large set of data pairs.
In both paradigms, $\phi_{0 \to 1}$ is modeled discretely and directionally, utilizing the volumetric or the spline-based representation. This representation implies that $\phi_{0 \to 1}: \mathbb{Z}^3 \to \mathbb{R}^3$ maps a coordinate $z_0$ on the grid (where $z_0 \in \mathbb{Z}^3$) of $I_0$ to a continuous location $x_1$ in $I_1$. 
The trilinear interpolation operator is required to calculate the displacement vector for a continuous location $x_0 \in \mathbb{R}^3$.

With the estimated DVF $\phi_{0 \to 1}$, we naturally have linear motion modeling. Built upon this, 4D frame interpolation estimates the intermediate frame $I_t, t\in (0, 1)$. In previous approaches, two directions have been pursued: 
\textbf{Forward Warping:} The DVF for the intermediate frame is approximated by $\phi_{0 \to t} \approx t \cdot \phi_{0 \to 1}$, and a trained neural network handles the challenging task of forward warping $I_0 \bullet \phi_{0 \to t}$. 
This model introduces significant errors: a) As shown in Figure.~\ref{fig:4.1a}, $\phi_{0 \to t} \approx t \cdot \phi_{0 \to 1}$ is not precise when the motion is nonlinear; b) Forward warping suffers from holes and multiple sources mapping~\cite{niklaus2020softmax}, as illustrated in Figure~\ref{fig:4.1c}. \textbf{Backward Warping:} Researchers adopt the backward warping $I_1 \circ \phi_{t \to 1}$, but this method suffers from inaccuracies in approximating $\phi_{t \to 1}$. As shown in Figure~\ref{fig:4.1b}, $\phi_{t \to 1} \neq (1-t) \cdot \phi_{0 \to 1}$, even for the linear motion. This discrepancy arises because the vector $(1-t) \cdot \phi_{0 \to 1}$ starts from a continuous location $x_t \in \mathbb{R}^3$, while $\phi_{t \to 1}$ for warping requires the vector to start from a discrete grid-point $z_t \in \mathbb{Z}^3$.

\subsection{Continuous sPtatial and Temporal Representation}

\subsubsection{Spatial Continuous Modeling}

\begin{figure}[ht]
    \centering
    \begin{minipage}[c]{0.48\textwidth}
        \centering
        \includegraphics[width=\textwidth]{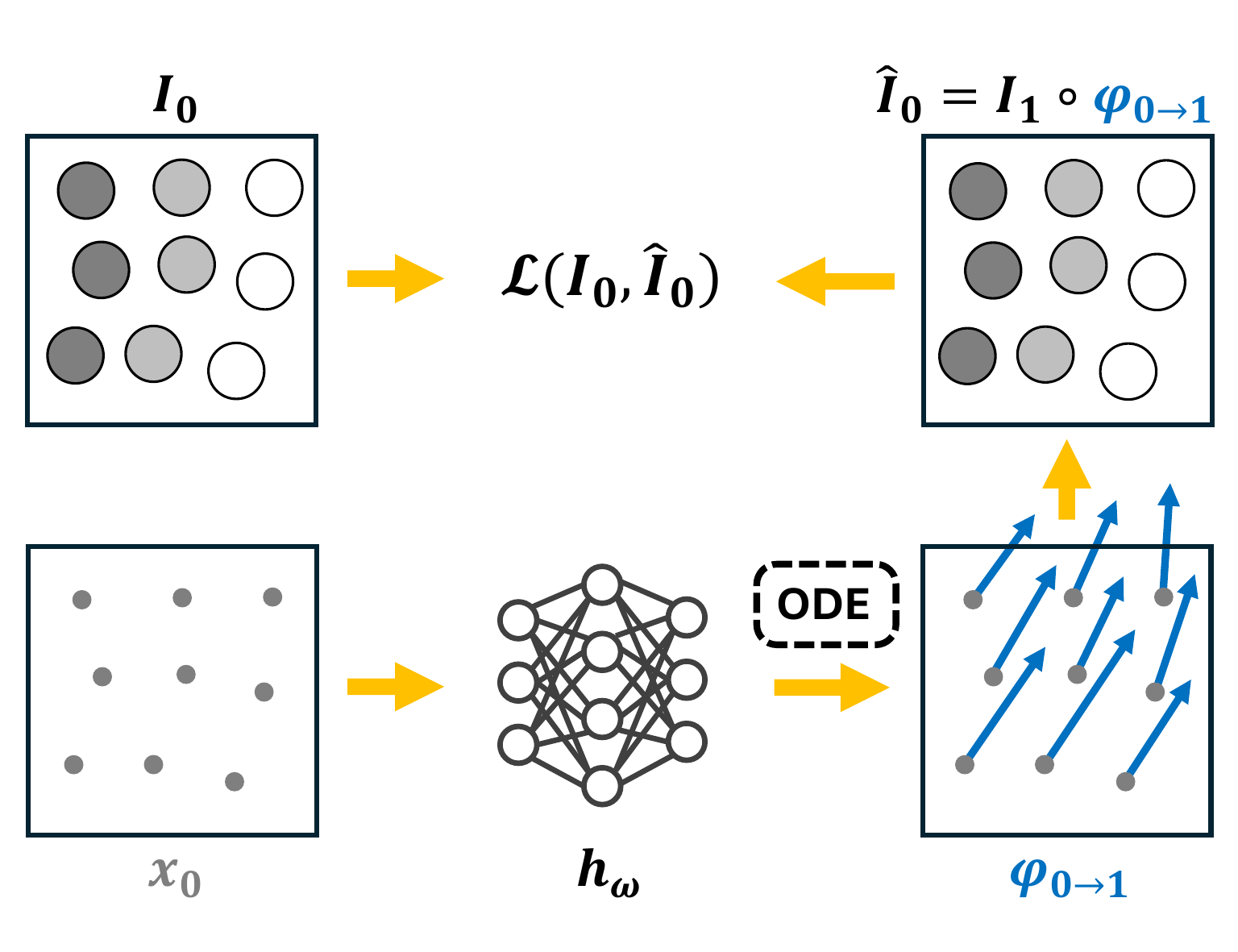}
        \subcaption{Siren network for spatial continuity.}
        \label{fig:4.2a}
    \end{minipage}
    \begin{minipage}[c]{0.48\textwidth}
        \centering
        \includegraphics[width=\textwidth]{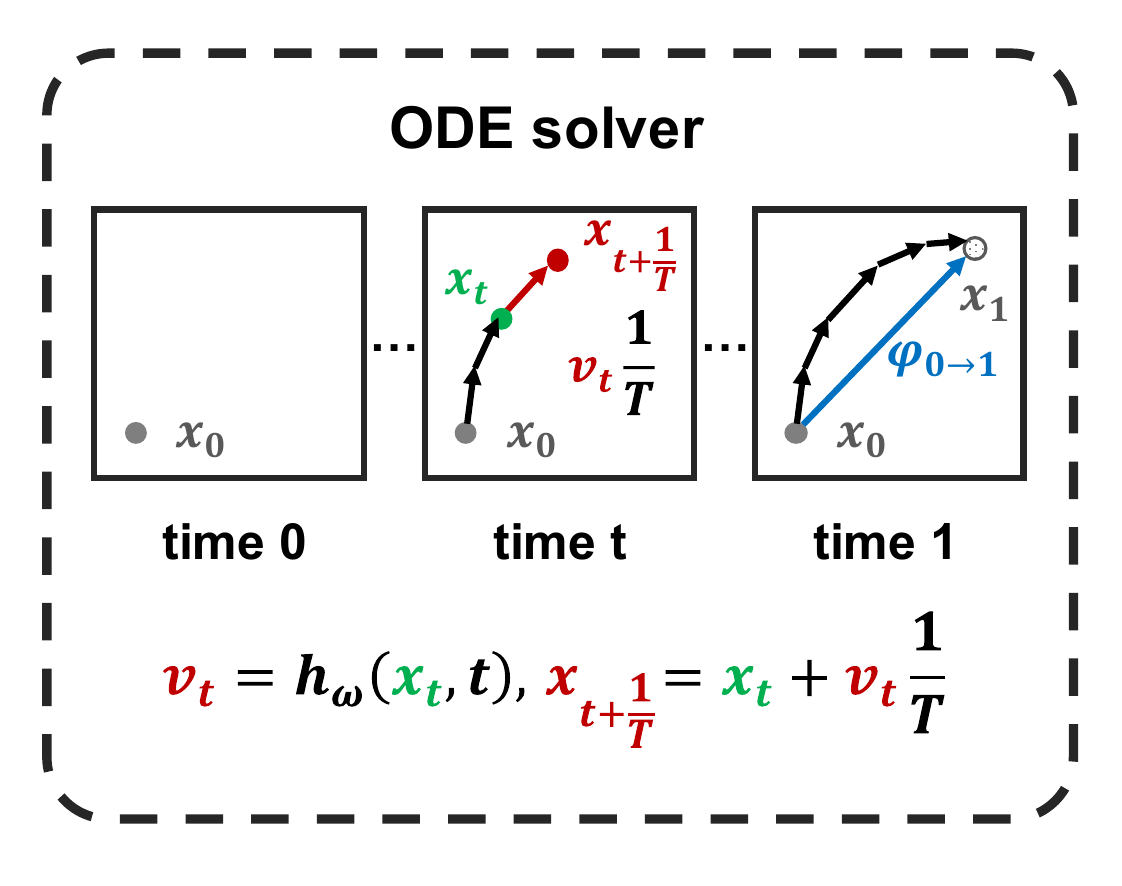}
        \subcaption{ODE solver for temporal continuity.}
        \label{fig:4.2b}
    \end{minipage}
    \caption{The proposed Continuous sPatial and Temporal (CPT) modeling. Given continuous input locations $x_0$, the Siren network $h_\omega$ estimates their corresponding velocities $\upsilon_t$ at time $t$. The full trajectory is then calculated by the ODE solver over $\upsilon_t, t\in [0, 1]$, yielding the displacement vector $\varphi_{0 \to 1}$ for optimizing the deformable image registration (DIR). The CPT motion modeling then naturally enables continuous image interpolation.}
    \label{fig:4.2}
    \vspace{5mm}
\end{figure}

The above dilemma faced in \textcolor{red}{\sout{deformable image registration}} DIR and video frame interpolation (VFI) estimation arises from the discrete modeling in both space and time. Only through the spatial continuous modeling can we precisely estimate the continuous DVF $\varphi _{0 \to 1}: \mathbb{R}^3 \to \mathbb{R}^3$.
To achieve this goal, we leverage the adaptivity of INR, which has achieved great success in 3D reconstruction. Instead of using INR to reconstruct the 3D image, we use it to model the displacement, resulting in $\varphi_{0 \to 1}(x) = g_{\theta}(x) + x$, where $g_{\theta}$ is implemented as a Siren\cite{sitzmann2020implicit} network, as in IDIR\cite{wolterink2022implicit}. The optimization target for DIR then becomes:
\begin{equation}
    \theta^* = \underset{\theta}{\arg\min} D(I_0, I_1 \circ \varphi_{0 \to 1}) + \lambda_1 R_1(\varphi_{0 \to 1}) + \lambda_2 R_2(\theta),
\end{equation}
where regularization is also applied to the network parameters $\theta$. The process is illustrated in Figure~\ref{fig:4.2a}, where the network is $g_\theta$ instead in this modeling.
This approach avoids the numerical accuracy sacrifice from trilinear interpolation, as above. Additionally, as both input and output are continuous coordinates, the DVF reversing approximations become feasible. In conclusion, spatial continuity eliminates the discrete and directional limitations of conventional representations. However, achieving continuous frame interpolation remains elusive, as accurately estimating $\varphi_{t \to 1}$ is still challenging.

\subsubsection{Temporal Continuous Modeling}\label{sec:temp-cont}

In Ordinary Differential Equation (ODE)-~\cite{ode1,ode2,ode3} and Large Deformation Diffeomorphic Metric Mapping (LDDMM)~\cite{lddmm}-based DIR methods, large deformations are decomposed into smaller steps to ease the burden of correspondence searching. Although we have not attempted to solve the problem of large deformations, we have found a solution for temporal continuity by decomposition. We estimate the velocity vector field (VVF) $\upsilon$ instead of the DVF to achieve this. The VVF can be estimated as a function of $t$ as $\upsilon_t(x) = h_{\omega}(x, t)$ or independent of $t$ as $\upsilon(x) = h_{\omega}(x)$, where $\omega$ is the parameter for the Siren network $h$. The DVF is then calculated from the temporal integration of the VVF:
\begin{equation}
    \varphi_{0 \to 1} (x_0) = x_0 + \int_0^1 h_{\omega}(x_{t}, t) dt,
\end{equation}
where $x_t$ represents the particle or tissue's location at time $t$. For practical implementation, this needs to be estimated discretely in time, using the Euler method:
\begin{equation}\label{eq:vvf}
    \varphi_{0 \to 1} (x_0) = x_0 + \sum_{t=0}^1 h_{\omega}(x_{t}, t) \frac{1}{T},
\end{equation}
where $T$ is the maximum number of steps for the ODE solver; typically, more steps yield less loss. As illustrated in Figure~\ref{fig:4.2b}, this can also be recursively calculated as:
\begin{equation}
    \varphi_{0 \to 1} (x_0) = x_1; \quad 
    x_{t + \frac{1}{T}} = x_{t} + h_{\omega}(x_{t}, t) \frac{1}{T}.
\end{equation}

Since the VVF can be reversed directly, we can easily estimate the reversed DVF $\phi_{1 \to 0}(x_1) = x_1 - \sum_{t=1}^0 h_{\omega}(x_{t}, t) \frac{1}{T}$. Thus, the optimization target for DIR becomes:
\begin{equation} \label{eq:loss}
\begin{split}
    \omega^* &= \underset{\omega}{\arg\min} D(I_0, I_1 \circ \varphi_{0 \to 1}) + \lambda_1 R_1(\varphi_{0 \to 1}) + \lambda_2 \sum_{t=0}^1 R_2(\upsilon_{t}) + \lambda_3 R_3(\omega) \\
    &+ D(I_1, I_0 \circ \varphi_{1 \to 0}) + \lambda_1 R_1(\varphi_{1 \to 0}),
\end{split}
\end{equation}
which also constrains the diffeomorphism property of the mapping.

Our proposed method distinguishes us from ODE-based and LDDMM-based methods, which still rely on an Eulerian specification that models only from the coordinates' perspective. These methods require discrete steps to integrate over time, involving successive warping of the VVF, such as $\phi_{0 \to 1} \approx \phi_{0 \to 0.5} \circ \phi_{0.5 \to 1}$. The more steps involved, the more trilinear interpolation is needed, decreasing precision. In contrast, our implementation is the first to seamlessly bridge the Eulerian and Lagrangian specifications, accommodating both the coordinates' and parcels' perspectives. The first perspective allows its easy formation as a mapping network, while the second provides the possibility for parcel/tissue tracking. 

\subsection{Extension for Frame Interpolation}

\begin{figure}[t!]
    \centering
    \begin{minipage}[c]{0.40\textwidth}
        \centering
        \includegraphics[width=1.0\textwidth]{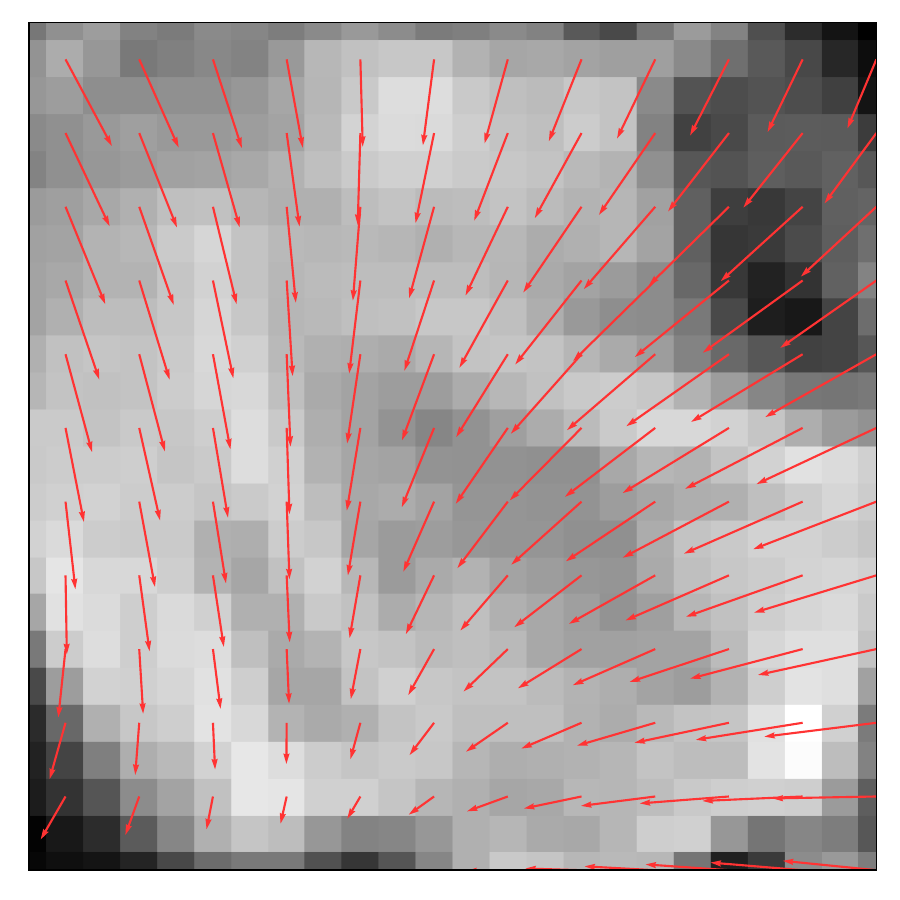}
        \subcaption{Pure discrete.}
        \label{fig:4.3a}
    \end{minipage}
    \begin{minipage}[c]{0.40\textwidth}
        \centering
        \includegraphics[width=1.0\textwidth]{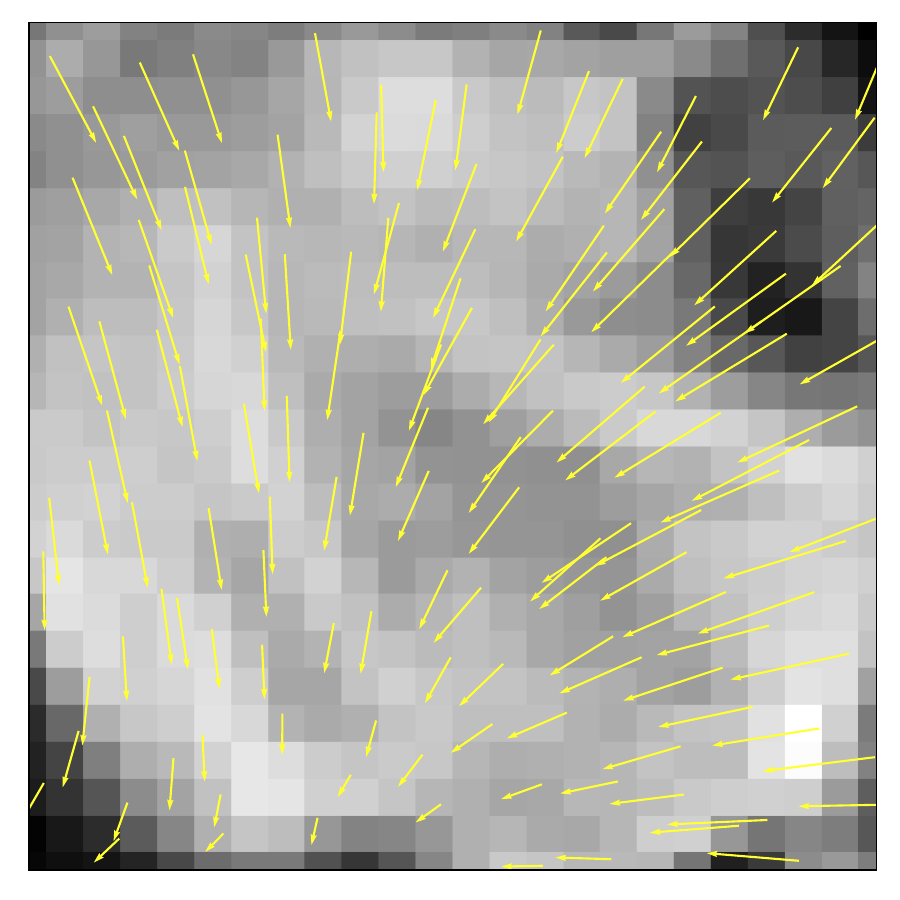}
        \subcaption{Spatial continuous.}
        \label{fig:4.3b}
    \end{minipage}
    \begin{minipage}[c]{0.40\textwidth}
        \centering
        \includegraphics[width=1.0\textwidth]{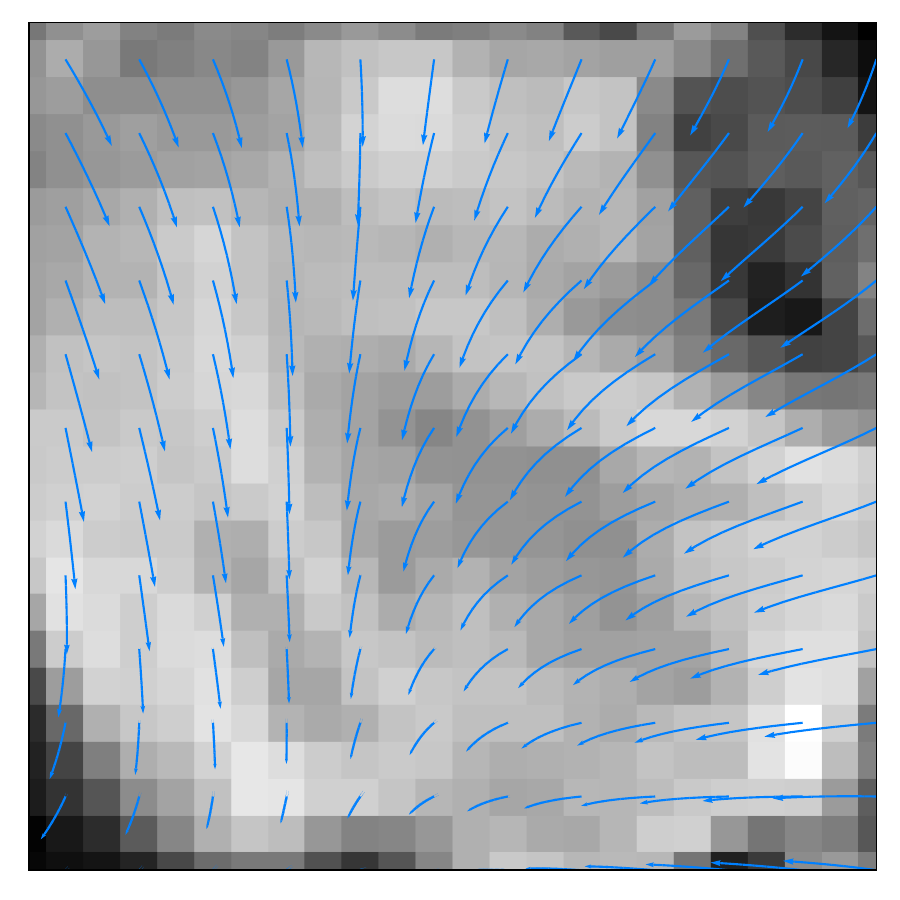}
        \subcaption{Temporal continuous.}
        \label{fig:4.3c}
    \end{minipage}
    \begin{minipage}[c]{0.40\textwidth}
        \centering
        \includegraphics[width=1.0\textwidth]{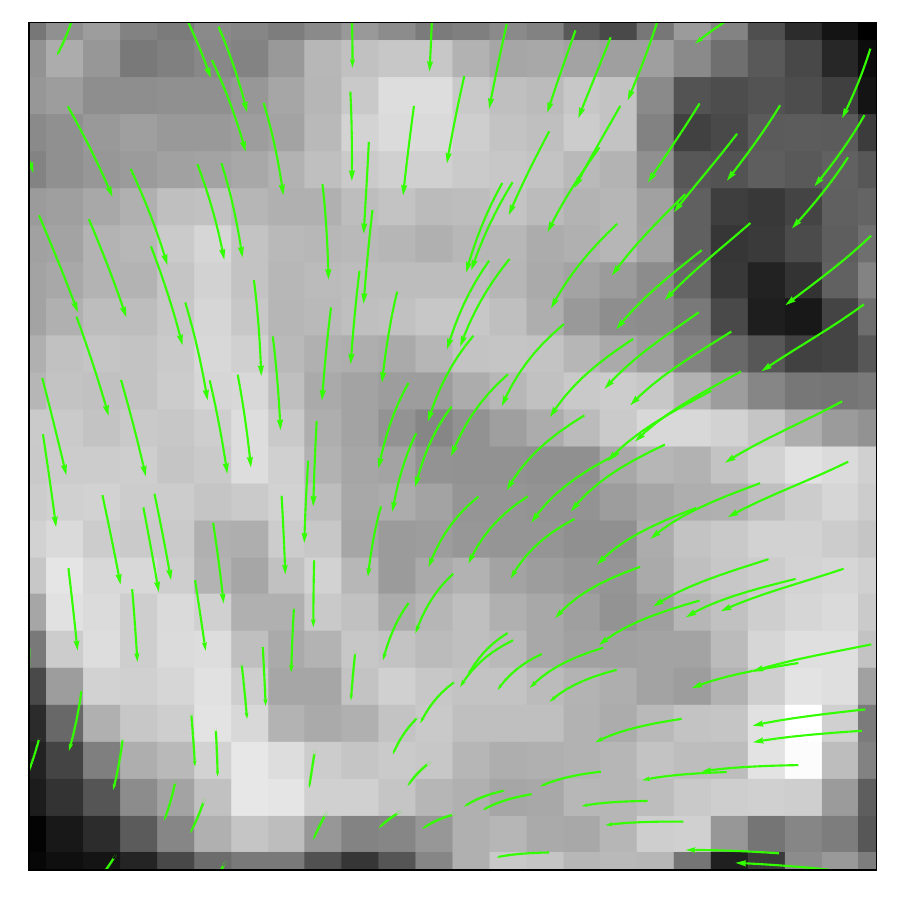}
        \subcaption{Both continuous.}
        \label{fig:4.3d}
    \end{minipage}
    \caption{Visualization of motion under different continuity settings. (a) In pure discrete modeling, the motion arrow can only start from a discrete location on the image grid and linearly point toward the target location. (b) Temporal continuity allows for smooth trajectories, encoding the dynamic of the motion. (c) The proposed Continuous sPatial and Temporal modelling (CPT) allows the motion to start from any continuous location while following a smooth curve.}
    \label{fig:4.3}
    \vspace{5mm}
\end{figure}

As illustrated in Figure~\ref{fig:4.3d}, the continuous spatial and temporal representation leads to smooth displacement curves, starting from anywhere inside the image. As a comparison, the temporal-only continuous model allows for a smooth trajectory but limits the starting points on the grid points $x_0 \in \mathcal{R}^3$, illustrated in Figure~\ref{fig:4.3c}. Besides, spatial-only contiguous models' trajectories can start anywhere but proceed directly as in Figure~\ref{fig:4.3b}. While the pure discrete can only model straight displacement vectors, shown in Figure~\ref{fig:4.3a}. Equipped with properties of spatial and temporal continuity, our motion model can derive the displacement vector for any time $t_0 \in (0, 1)$ to any time $t_1 \in (0, 1)$ from any location $x_0$ using an ODE solver (such as the Euler method~\cite{hildebrand1987introduction}):
\begin{equation}
    \varphi_{t_0 \to t_1}(x_{t_0}) \approx x_{t_0} + \sum_{t=t_0}^{t_1} \upsilon_t \frac{1}{T},
\end{equation}
where the loss of numerical precision only comes from $\delta t$.

These formulations enable us to estimate $\varphi_{t \to 0}$ and $\varphi_{t \to 1}$ more precisely than previous methods. We avoid the cumbersome forward warping because we don't utilize $\varphi_{0 \to t}$ and $\varphi_{1 \to t}$. Instead, we calculate frame interpolation at time $t$ using the weighted average of $I_1 \circ \varphi_{t \to 1}$ and $I_0 \circ \varphi_{t \to 0}$ as:
\begin{equation}
    I_t = w_1 \cdot I_1 \circ \varphi_{t \to 1} + w_0 \cdot I_0 \circ \varphi_{t \to 0}.
    \label{eq:interp}
\end{equation}

The naive approach would simply \textbf{average} the weights ($w_0$ and $w_1$). However, we choose to adopt \textbf{linear} weights ($w_1 = t, w_0 = (1 -t)$). It is worth noting that in our approach, no post-processing or so-called refinement network is adopted.

\subsection{Dataset and Evaluation Metrics}

This study utilizes the openly available DIR-Lab dataset~\cite{castillo2009framework}, tailored for research in DIR algorithm development. The dataset comprises 4DCT scans from 10 patients undergoing treatment for esophageal malignancies. Each patient dataset consists of 10 respiratory phases, spanning from $0\%$ to $90\%$ of the respiratory cycle in increments of $10\%$. The CT images have been pre-processed to include 300 manually identified landmark points for the two extreme phases ($0\%$ and $50\%$), facilitating the evaluation of the landmark accuracy. Following the dataset’s convention, the $0\%$ phase (end of inhalation) CT is selected as the fixed image, while the $50\%$ phase (end of exhalation) is chosen as the moving image. In addition to landmark points, a series of organs at risk (OARs) were extracted using Totalsegmentor~\cite{wasserthal2023totalsegmentator} and reviewed and revised by physicians to ensure adherence to clinical standards.
For the interpolation task, the intermediate phases ($10\%$ to $40\%$) are used as the ground truth to evaluate the performance.
Except for Mean Absolute Error (MAE)~\cite{hodson2022root}, we include Peak Signal-to-Noise Ratio (PSNR)~\cite{johnson2006signal}, NCC, Structural Similarity Index Measure (SSIM)~\cite{brunet2011mathematical}, and Normalized Mean Squared Error (NMSE) to evaluate the frame interpolation.

\subsection{Implementation Specifications}

\subsubsection{Specifications for Deformable Image Registration}

The CPT-DIR algorithm was implemented into two specific models for validation and comparison: (a) the spatial continuous-only method as end-to-end (denoted as E2E) employing a regularization term $R_1$ solely as the Jacobian determinant~\cite{wolterink2022implicit} over the DVF; (b) the combined temporal and spatial continuous method as large deformation decomposition (denoted as LDD) utilizing the regularization term $R_2$ only composed of the L2 norm of the VVF to avoid Zigzag trajectory. Both models were trained with the NCC loss and subsequently underwent two evaluation scenarios. In the first scenario, we followed the exact guidelines of the DIR-lab dataset to provide a fair comparison to previously submitted algorithms. The training was confined to the lung region only. Evaluation metrics in this scenario included Target Registration Error (TRE)~\cite{datteri2012estimation} for landmark accuracy, \textcolor{red}{\sout{Mean Absolute Error}}MAE for image similarity in the lung region, and negative Jacobian determinant ratio of DVFs (Neg JacDet) for diffeomorphism. In the second scenario, we evaluated the DIR registration performance for the RT dose accumulation application. The training was conducted on the entire body to assess the sliding boundary region (ribcage) performance fully. No auto- or manual segmentation was applied in this scenario. Evaluation criteria in this scenario included the landmark accuracy TRE, and whole body MAE first. Then, the Dice coefficient and MAE were explicitly calculated for each OAR (segmented by TotalSegmentor\footnote{\url{https://totalsegmentator.com/}}), including lung (left and right), esophagus, vertebra, heart, spinal cord, and ribcage. We systematically compared the new CPT-DIR's performance against the performance of the classic B-Splines algorithm (implemented in Plastimatch~\cite{sharp2010plastimatch}) and the value of the corresponding metric for a scenario without registration. 

\subsubsection{Specifications for 4D Frame Interpolation}

We implement $h_{\omega}$ as a 4-layer Siren network~\cite{sitzmann2020implicit}, with the hyper-parameter $\omega_0$ set to $48$ and the layer width set to $256$. The output is multiplied by the size of the voxel to guarantee the initial estimation of the $\varphi$ is within [-voxel, voxel]. The $T$ used in the ODE solver is set to the number of intermediate frames plus 1, and $\beta$ used in weighted merging is set to $4$. We linearly warm up the learning rate for the first 500 steps, adopt a cosine learning rate scheduler to stabilize the convergence, and train the network for 3,000 steps, each with a mini-batch of 10,000 voxels. To speed up optimization, we only used the regularization term $R_2$ in Equation~\ref{eq:loss} as $\sum_t \|\upsilon_t\|^2$ and neglected $R_1$ and $R_3$, with $\lambda_2$ set as 1e-4. We implement CPT-Interp with Pytorch on an NVIDIA RTX 4090 GPU and compare with previous methods for the 4D medical image interpolation, including the fully discrete conventional approach ANTs~\cite{avants2009advanced} and DL-based approach UVI-Net~\cite{kim2024data}, spatial continuous-only method IDIR~\cite{wolterink2022implicit}, temporal continuous-only methods ORRN~\cite{liang2023orrn} and NODEO~\cite{wu2022nodeo}. ORRN and UVI-Net are trained over the 4DLung dataset~\cite{4dlung}, while all the others are case-specific optimization methods without training. In this study, the ANTs registration framework was implemented using the TimeVaryingVelocityField transformation. Optimization was guided by minimizing a composite objective function comprising Cross-Correlation (CC) and Mean Squared Error (MSE) terms. For the IDIR method, regularization of the deformation field was achieved through the bending energy penalty. All remaining parameters for both were set to their respective default values as provided in their implementations.

%% file: sections-r1/3.results.tex
\section{Results}
\subsection{Results for Deformable Image Registration}
\begin{figure}[t!]
    \centering
    \includegraphics[width=\linewidth]{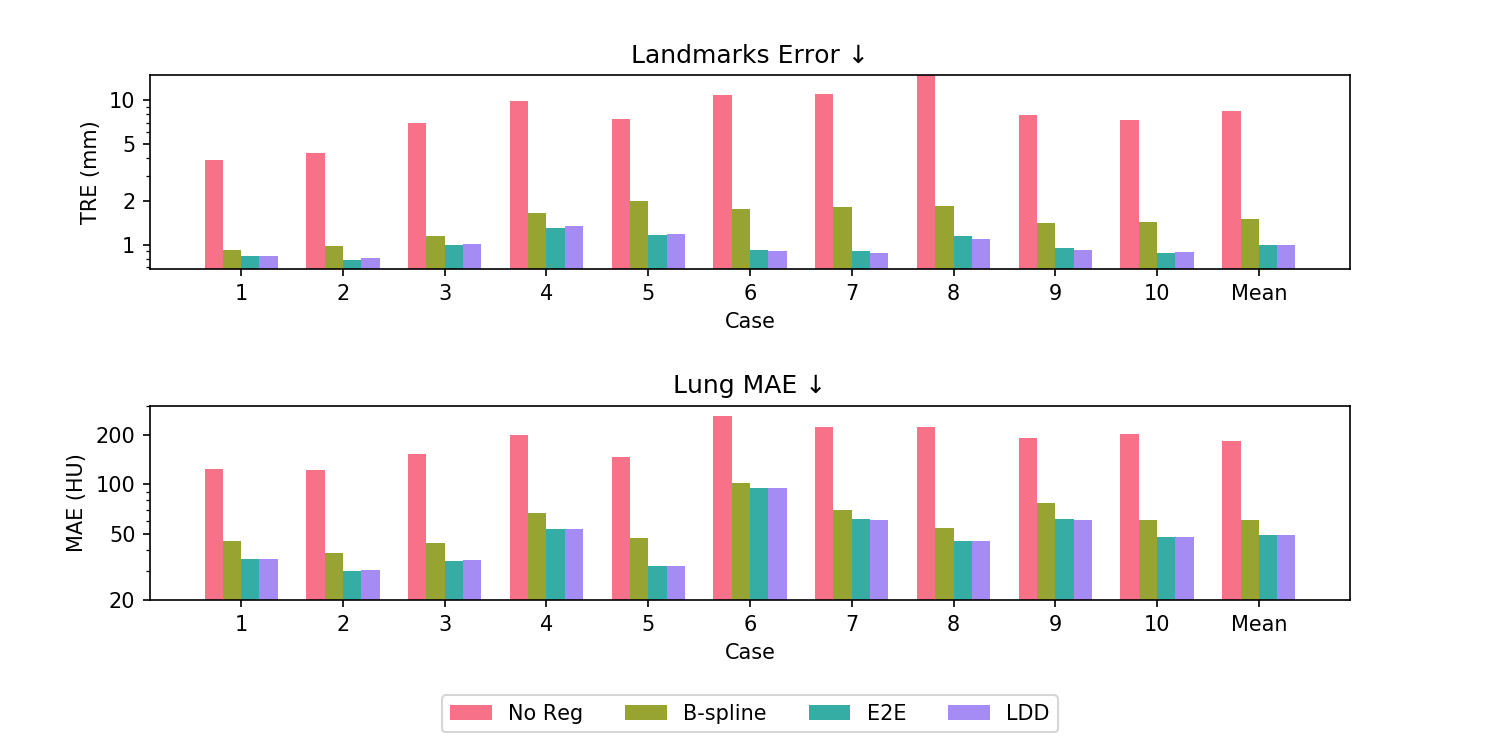}
    \caption{Quantitative comparisons of landmark accuracy and regional MAE between CPT-DIRs and classic B-spline based DIR algorithm. For fairness, the training was conducted inside the lung region.}
    \label{fig:4.4}
    \vspace{5mm}
\end{figure}
\begin{figure}
    \centering
    \includegraphics[width=0.8\linewidth]{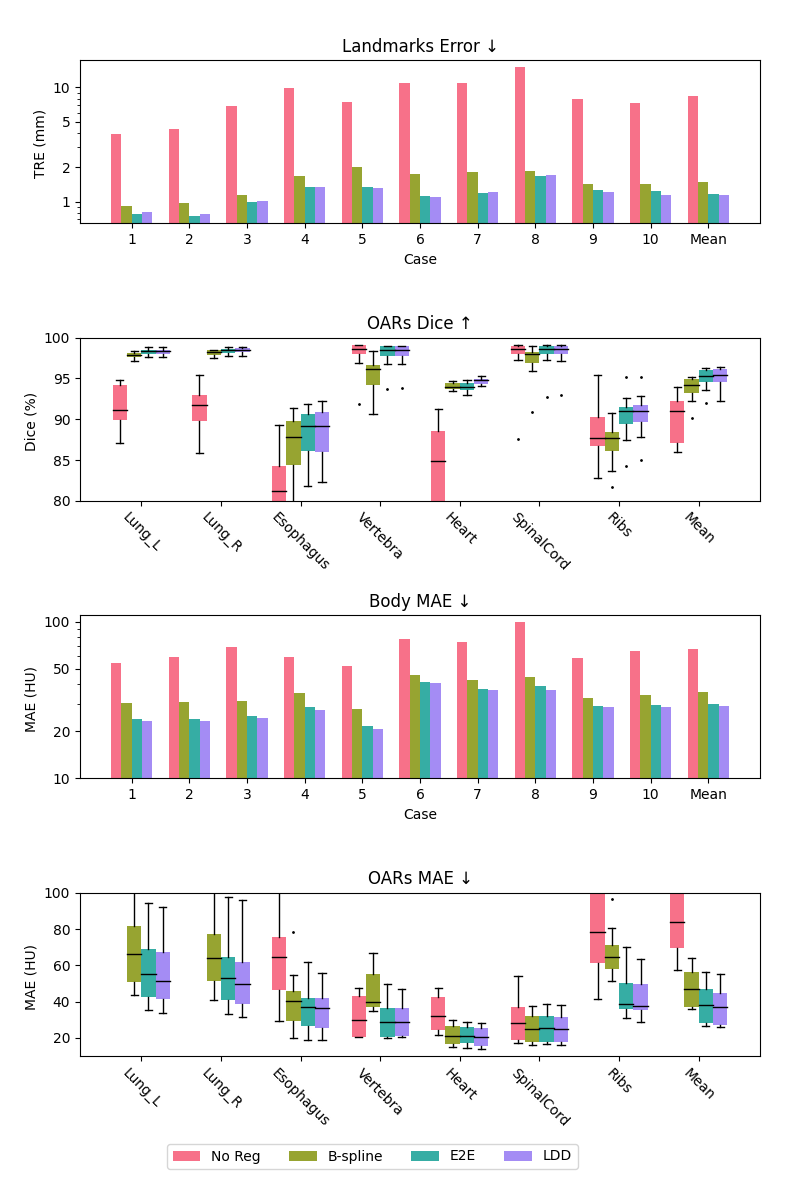}
    \caption{Quantitative comparison of landmark accuracy, regional MAE and OARs propagation accuracy between CPT-DIRs and classic B-spline based DIR algorithm. The training was conducted inside the whole-body region.}
    \label{fig:4.5}
    \vspace{5mm}
\end{figure}
\begin{figure}
    \centering
    \includegraphics[width=0.95\linewidth]{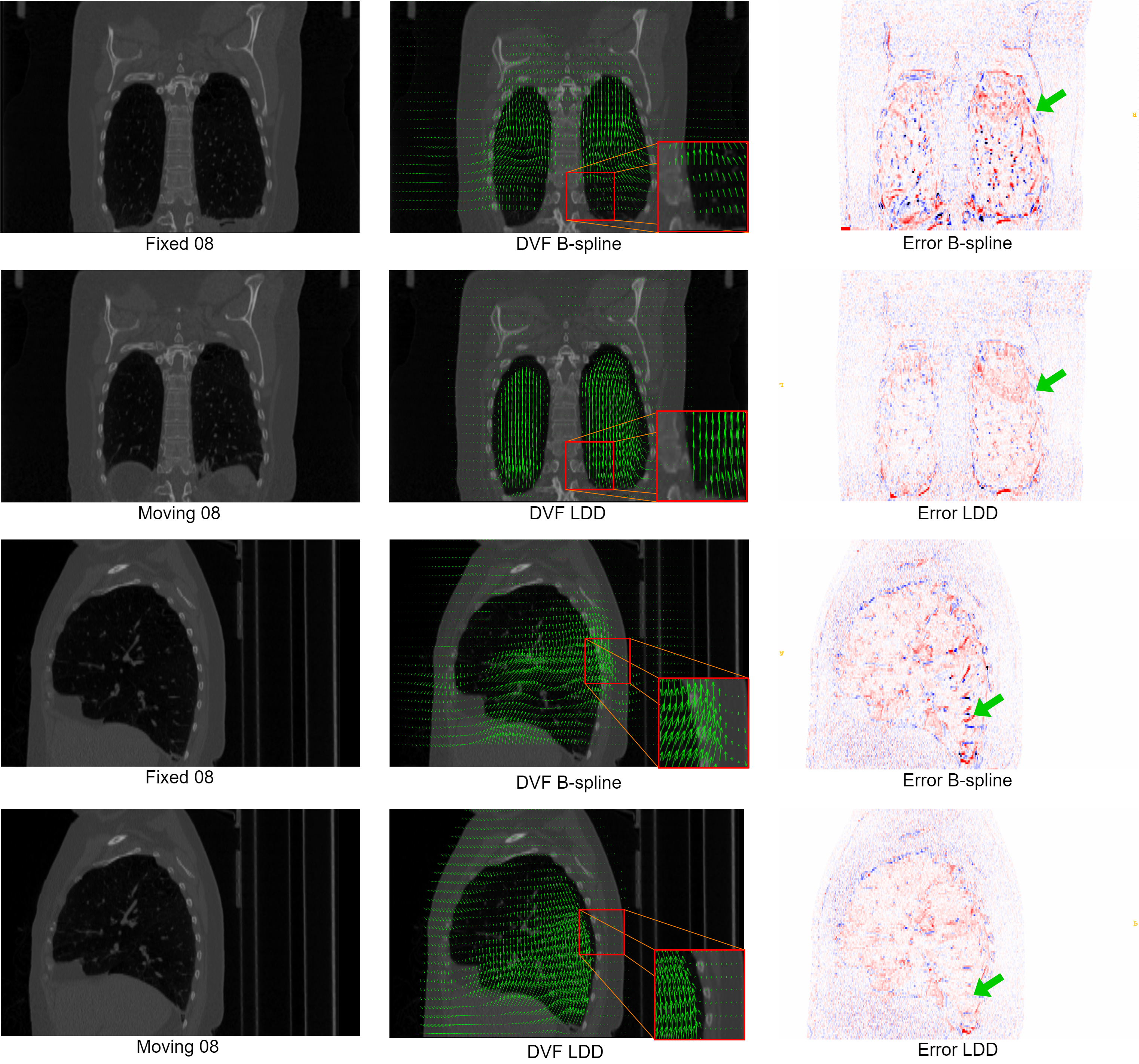}
    \caption{Visual comparison with B-spline over estimated DVFs and error maps. Arrows indicate the sliding boundary region.}
    \label{fig:4.6}
    \vspace{5mm}
\end{figure}
When training within the lung region alone, the landmark TRE for E2E and LDD models were recorded at $0.99 \pm 1.11$ mm and $0.99\pm1.07$ mm, respectively, significantly reducing B-spline's TRE of $2.79 \pm 1.88$ mm by more than half. By comparison, the original IDIR implementation~\cite{wolterink2022implicit} only achieved a TRE of $1.07 \pm 1.11$ mm. The leading state-of-the-art conventional method, iso PTV~\cite{vishnevskiy2016isotropic}, achieved a TRE of $0.95$ mm within 3 minutes, whereas CPT-DIR can achieve similar performance within only $15$ seconds on an RTX 4090 GPU. Regarding lung MAE, both E2E and LDD models reduced B-spline’s results from $60.75 \pm 59.06$ HU to $49.72 \pm 39.56$ HU and $49.79 \pm 39.73$ HU, respectively. Besides, LDD reduced the average negative Jacobian determinant ratio from $5e-5$ (B-spline) and $9e-5$ (E2E) to $0.00$. Detailed per-case results are shown in Figure~\ref{fig:4.4}.

When training within the entire body, the whole-body MAE, both E2E ($29.80 \pm 34.09$ HU) and LDD ($28.99 \pm 32.70$ HU) still considerably surpassed B-spline ($35.46 \pm 46.99$ HU). The landmark accuracy degraded slightly for both the E2E and LDD models to $1.17 \pm 1.36$ mm and $1.17 \pm 1.45$ mm, respectively, yet remained lower than half of the B-spline’s performance. Detailed per-case TRE, MAE, OAR-specific Dice, and MAE are shown in Figure~\ref{fig:4.5}. Among all OARs, ribs are vital regions for assessing registration quality on sliding boundaries. Without registration, the MAE and Dice coefficients for ribs stood at $75.40$ HU and $89.30\%$, respectively. B-spline offered only marginal improvement, with metrics at $65.65$ HU and $90.41\%$. In contrast, E2E and LDD models dramatically reduced ribcages MAE to $43.62$ HU and $42.04$ HU, respectively, while increasing Dice coefficients to $90.34\%$ and $90.56\%$. Notably, while most metrics show that LDD has a slight edge over E2E, the versatility of LDD is further emphasized by its landmark error of $1.17 \pm 1.15$ mm when executing a reverse trajectory.

Regarding computational efficiency, B-spline took an average of $65.34 \pm 35.43$s, while E2E and LDD processed at faster rates of $11.96 \pm 1.68$s and $14.32 \pm 2.34$s, respectively. Additionally, the visualizations of the resulting DVFs and error maps (between warped and fixed images) from different DIR methods can be found in Figure~\ref{fig:4.6}. As an example, case-04 (the largest landmark TRE) and case-08 (the largest motion) show that the new CPT-DIR methods can reduce the difference in the ribcage region due to the improved capture of sliding boundary motion.

\subsection{Results for Frame Interpolation}

As shown in Table~\ref{tab:interp}, CPT-Interp outperforms all previous methods across all metrics. 
Noteworthy, CPT-Interp requires no training dataset and only needs only unsupervised testing time optimization (TTO) over the test cases. This means CPT-Interp can generalize freely to different datasets without fine-tuning or domain adaptation. 
Additionally, CPT-Interp requires no further post-processing or refinement network. Moreover, the TTO only takes $1.96$s,
and inference for all intermediate frames takes approximately $1.23$s, which is much faster than the instance-specific optimization time in UVI-Net ($70$ s) and TTO in IDIR (1 min).

\begin{table}[t!]
    \centering
    \caption{Quantitative comparison of frame interpolation results. All metrics are averaged over all intermediate frames and repeated three times.}
    \vspace{1mm}
    \small
    \scriptsize
    \tabcolsep=3.0mm
    \begin{tabular}{l|cccccc}
    \toprule
    Method                        & TRE $\downarrow$ & MAE $\downarrow$& PSNR $\uparrow$       & NCC $\uparrow$            & SSIM $\uparrow$           & NMSE $\downarrow$         \\ \midrule
    ANTs~\cite{avants2009advanced}& $2.61\pm 0.90$ & $40.50\pm6.72$ & $32.97\pm2.36$         & $0.541\pm0.102$           & $0.895\pm0.028$           & $1.849\pm0.539$           \\
    ORRN~\cite{liang2023orrn}     & $2.40\pm0.75$ & $23.51\pm3.50$ & $38.30\pm1.07$         & $0.691\pm0.085$           & $0.950\pm0.012$           & $0.576\pm0.258$           \\
    UVI-Net~\cite{kim2024data}      & $1.97\pm0.39$ & $21.02\pm4.13$ & $39.66\pm1.37$         & $0.708\pm0.078$           & $0.955\pm0.013$           & $0.455\pm0.277$           \\
    NODEO~\cite{wu2022nodeo}    & $1.92\pm0.41$ & $33.65\pm5.72$ & $34.35\pm2.95$         & $0.588\pm0.103$           & $0.926\pm0.022$           & $1.371\pm0.502$           \\
    IDIR~\cite{wolterink2022implicit}      & $1.90\pm0.45$ & $18.93\pm3.90$ & $39.76\pm1.48$         & $0.717\pm0.044$           & $0.960\pm0.011$           & $0.456\pm0.278$           \\
    CPT-Interp                    & $\mathbf{1.87\pm0.39}$ & $\mathbf{17.88\pm3.79}$ & $\mathbf{40.26\pm1.58}$& $\mathbf{0.731\pm0.045}$  & $\mathbf{0.964\pm0.011}$  & $\mathbf{0.414\pm0.271}$  \\ \bottomrule
    \end{tabular}
    \label{tab:interp}
    \vspace{5mm}
\end{table}

\begin{figure}[ht]
    \centering
    \includegraphics[width=1.0\textwidth]{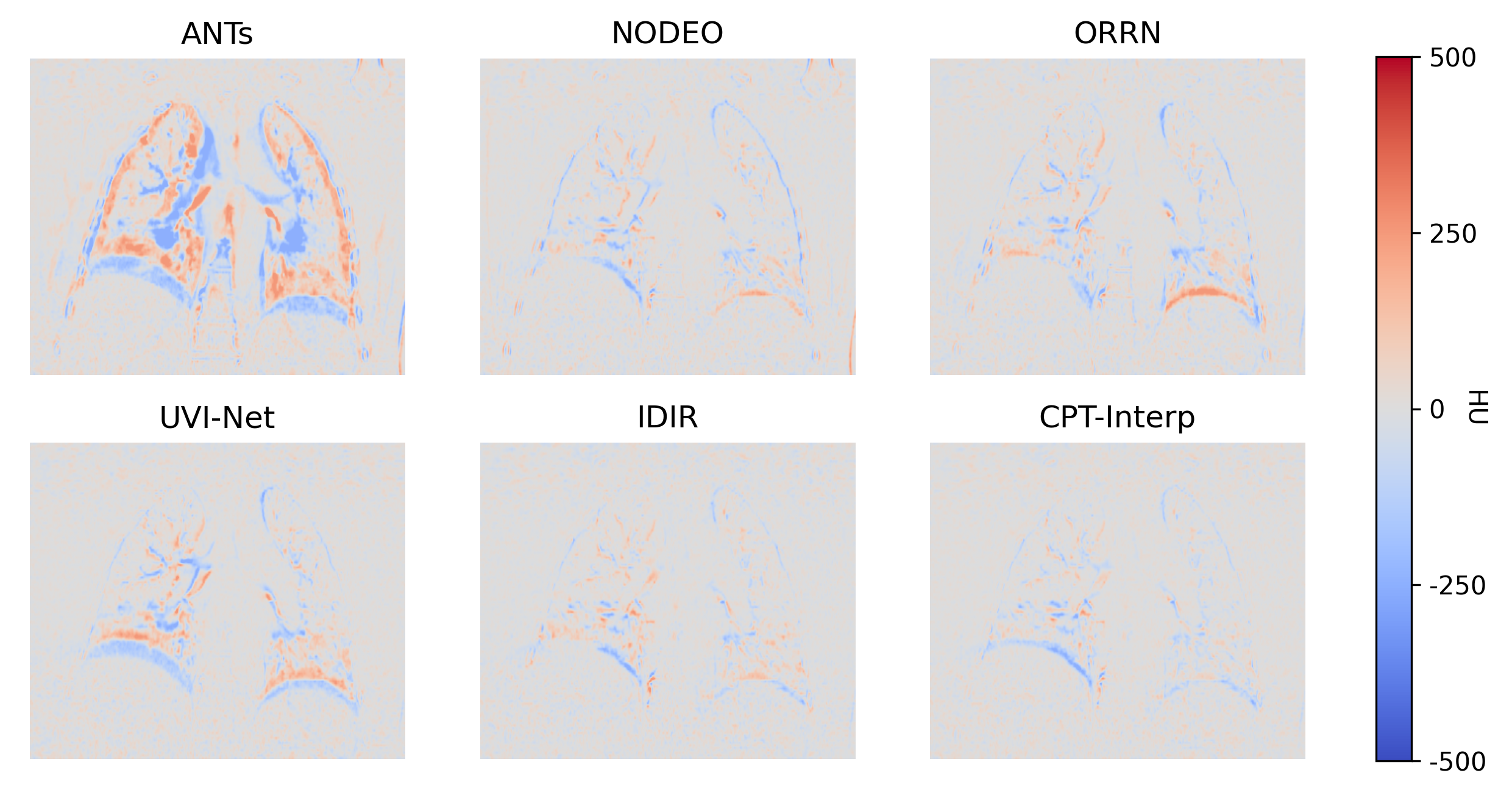}
    \caption{Visualization of the error maps for the interpolated frames over the DIRLab dataset. The error map is calculated as the interpolated image minus the ground-truth image.}
    \label{fig:4.8}
    \vspace{5mm}
\end{figure}

To qualitatively evaluate the performance of our CPT-Interp method, we visualize the error maps for the interpolated frames compared to the ground truth in Figure~\ref{fig:4.8}.
As illustrated, CPT-Interp consistently produces error maps with fewer high-error regions compared to the other methods, indicating a closer approximation to the ground truth.

%% file: sections-r1/4.discuss.tex
\section{Discussion}
To enhance DIR performance for RT applications, we draw inspiration from IDIR~\cite{wolterink2022implicit} and LDDMM by incorporating spatial and temporal continuity into our LDD models. We aimed to improve intra-fraction motion modeling closer to the nature of physical anatomy. As elaborated in the introduction, the two prevailing learning paradigms for solving DIR problems are optimization-based and DL-based approaches. While optimization-based methods offer personalized modeling accuracy, they are time-consuming due to per-case parameter optimization. In contrast, DL-based methods leverage large datasets for training and provide real-time inference but may sacrifice performance. The newly proposed CPT-DIR and CPT-Interp methods here are also optimized per case without individual setting parameter optimization, aligning well with the former paradigm and offering personalized enhanced accuracy. This also satisfies the nature of radiotherapy, where a large dataset is challenging to collect, clean, or align the formats. Unlike previous optimization-based methods, our approach benefits from the computational efficiency of neural networks and GPU (benefits from DL-based methods), reducing processing time from minutes to seconds.

The DIR problem is inherently ill-posed and lacks a definitive ground truth, necessitating the need to parameterize the search space. Parameterization serves as a form of implicit regularisation, effectively reducing the space in the solution. Voxel-based methods address this challenge by optimizing each voxel independently, often necessitating additional regularisation terms to constrain voxel relationships. Namely, DL-based methods also operate in a voxel-based fashion. On the other hand, B-spline methods reduce the parametrization space from individual voxels to splines, leading to overly smooth results and difficulty in effectively handling sliding boundaries. In another direction, advanced regularizations have been proposed to solve the sliding boundary issue directly~\cite{fu2018adaptive}. In contrast, the new continuous motion modeling approach proposed here harnesses the power of neural networks to determine the smoothness/sharpness of each area automatically. Consequently, it can simultaneously model the smooth regions inside OARs and the sharp boundaries of the nearby organs.

Another significant challenge in DIR is the management of large deformations. Conventional methods like Demons rely on local assumptions, rendering them ineffective when faced with these conditions. This issue exacerbates the search space for matching locations in other DIR methods, and significant motion may disrupt local regularisation terms, affecting optimization targets. LDDMM and ODE-based methods~\cite{wu2022nodeo,liang2023orrn} address this challenge through decomposition, leveraging the typically more straightforward structure of the VVF compared to the DVF. The effectiveness of modeling large deformations through temporal continuous modeling extends beyond motion estimation, encompassing other DIR tasks in radiotherapy, such as inter-fractional anatomic changes. 

However, previous ODE-based methods and LDDMM were limited by their reliance on voxel-based representations, rooted in the Euler specification from the theory of fluid mechanics. Despite being designed under the Lagrange specification to track each parcel separately, their implementation was confined to the Euler specification. Our spatial continuous modeling approach estimates the VVF at any point within the image and at any time between two frames, not limited to grid points. Consequently, our implementation fully adheres to the Lagrange specification, enabling the tracking of each tissue component individually. 

Another advantage of temporal continuous modeling is the ease of reversing the DVF~\cite{murr2023applicability}. Conventional methods typically require double the effort to estimate bidirectional DVFs. However, in our framework, since the VVF can be directly reversed, reversing the DVF becomes straightforward. This capability enables tracking from any time between two frames and from any location inside the image to another time and location with just one-time training. This facilitates super-frame rate imaging, potentially improving 4DCT reconstruction algorithms and 4D imaging pipelines in the future.

Quality assurance is crucial for commissioning DIR methods, especially for clinical applications. Landmarks provided by DIR-Lab serve as a reliable benchmark, but their accurate labeling requires substantial human efforts. Alternatively, organ masks can be generated using open-source segmentation tools like Totalsegmentor or other commercial tools, enabling rapid validation of fixed and moving images. In the absence of landmarks or organ masks, image similarity metrics can serve as a validation measure, albeit primarily effective in boundary regions. Moreover, robustness is another critical consideration when evaluating DIR performance. The original IDIR implementation~\cite{molaei2023implicit} encountered higher failure rates in some cases in DIR-Lab, attributed to initialization schemes. We addressed this issue by adjusting initial estimations within a two-voxel size range. Besides, the application of our proposed method in clinical pipelines requires the estimation of uncertainty, which represents the next step in our ongoing work.


%% file: sections-r1/5.conclusion.tex
\section{Conclusion}
In summary, this work introduces a novel approach, CPT-DIR, that leverages continuous spatial and temporal modeling for DIR. By integrating principles of INR and LDDMM, CPT-DIR effectively handles sliding organ boundaries and significant anatomical changes over time, overcoming the limitations of traditional voxel-based and discrete motion modeling approaches. The tangible benefits of CPT-DIR are evident in its superior performance and efficiency compared to classic B-spline methods, as well as its enhanced adaptability and robustness in radiotherapy applications. Furthermore, we demonstrate the extensibility of CPT-DIR by applying it to the task of 4D medical image interpolation, referred to as CPT-Interp. It achieves superior accuracy and speed in interpolating medical images without requiring extensive training datasets or fine-tuning. The successful application of CPT-DIR to both image registration and interpolation tasks highlights its potential as a unified framework for advancing medical image analysis.

%% file: sections-r1/6.appendix.tex
\section{Extra Benchmark}

Beyond the DIR-Lab dataset, we further evaluated the proposed methods using the recently published Abdominal-DIR-QA dataset~\cite{criscuolo2025vessel}. This dataset was specifically designed to facilitate the development and validation of deformable image registration (DIR) algorithms for abdominal CT scans. It consists of 30 pairs of abdominal CT images, sourced from various publicly available repositories and the authors' institution. Each pair comprises scans from the same patient acquired on different days, presenting a challenging registration scenario due to significant organ deformations and inconsistent image content. The dataset includes a total of 1,895 highly accurate landmark pairs on matching blood vessel bifurcations, with an average of 63 landmark pairs per case. For our evaluation, cases 5 and 16 were excluded due to extreme field-of-view differences between the fixed and moving images, which prevented successful registration by both methods. In addition to Target Registration Error (TRE) calculated using these landmarks, we also assessed the negative ratio of the Jacobian determinant for the derived deformation vector fields (DVFs). The comparative results, detailed in Table~\ref{tab:abd-dir}, demonstrate that LDD achieved substantially superior performance compared to both B-spline and E2E methods.

\begin{table}[ht]
\centering
\vspace{5mm}
\caption{Quantitative comparisons of landmark accuracy (TRE) and negative Jacobian determinant ratio (Neg JacDet) between CPT-DIRs and classic B-spline based DIR algorithm.}
\begin{tabular}{c|cccc}
\toprule
Methods                    & No Reg             & B-spline      & E2E            & LDD           \\ \midrule
TRE (mm) $\downarrow$        & $30.80\pm 6.55$  & $8.61\pm 7.92$& $5.90\pm 7.60$ & $\mathbf{4.79\pm 6.28}$ \\
Neg JacDet (\%) $\downarrow$ & $0.00\pm 0.00$   & $2.83\pm2.16$ & $1.56\pm 1.00$ & $\mathbf{0.03\pm 0.08}$ \\ \bottomrule
\end{tabular}\label{tab:abd-dir}
\end{table}

%% file: revision1.bbl
\begin{thebibliography}{10}

\bibitem{verhey1995immobilizing}
L.~J. Verhey,
\newblock Immobilizing and positioning patients for radiotherapy,
\newblock in {\em Seminars in radiation oncology}, volume~5, pages 100--114, Elsevier, 1995.

\bibitem{oh2017deformable}
S.~Oh and S.~Kim,
\newblock Deformable image registration in radiation therapy,
\newblock Radiation oncology journal {\bf 35}, 101 (2017).

\bibitem{nenoff2023review}
L.~Nenoff, F.~Amstutz, M.~Murr, B.~Archibald-Heeren, M.~Fusella, M.~Hussein, W.~Lechner, Y.~Zhang, G.~Sharp, and E.~V. Osorio,
\newblock Review and recommendations on deformable image registration uncertainties for radiotherapy applications,
\newblock Physics in Medicine \& Biology {\bf 68}, 24TR01 (2023).

\bibitem{rigaud2019deformable}
B.~Rigaud, A.~Simon, J.~Castelli, C.~Lafond, O.~Acosta, P.~Haigron, G.~Cazoulat, and R.~de~Crevoisier,
\newblock Deformable image registration for radiation therapy: principle, methods, applications and evaluation,
\newblock Acta Oncologica {\bf 58}, 1225--1237 (2019).

\bibitem{zhang2012respiratory}
Y.~Zhang, D.~Boye, C.~Tanner, A.~J. Lomax, and A.~Knopf,
\newblock Respiratory liver motion estimation and its effect on scanned proton beam therapy,
\newblock Physics in Medicine \& Biology {\bf 57}, 1779 (2012).

\bibitem{mok2020large}
T.~C. Mok and A.~C. Chung,
\newblock Large deformation diffeomorphic image registration with laplacian pyramid networks,
\newblock in {\em Medical Image Computing and Computer Assisted Intervention--MICCAI 2020: 23rd International Conference, Lima, Peru, October 4--8, 2020, Proceedings, Part III 23}, pages 211--221, Springer, 2020.

\bibitem{amstutz2024quantification}
F.~Amstutz, P.~G. D’Almeida, X.~Wu, F.~Albertini, B.~Bachtiary, D.~C. Weber, J.~Unkelbach, A.~J. Lomax, and Y.~Zhang,
\newblock Quantification of deformable image registration uncertainties for dose accumulation on head and neck cancer proton treatments,
\newblock Physica Medica {\bf 122}, 103386 (2024).

\bibitem{fischer2008ill}
B.~Fischer and J.~Modersitzki,
\newblock Ill-posed medicine—an introduction to image registration,
\newblock Inverse problems {\bf 24}, 034008 (2008).

\bibitem{sotiras2013deformable}
A.~Sotiras, C.~Davatzikos, and N.~Paragios,
\newblock Deformable medical image registration: A survey,
\newblock TIP {\bf 32}, 1153--1190 (2013).

\bibitem{ostergaard2008acceleration}
K.~{\O}stergaard~Noe, B.~D. De~Senneville, U.~V. Elstr{\o}m, K.~Tanderup, and T.~S. S{\o}rensen,
\newblock Acceleration and validation of optical flow based deformable registration for image-guided radiotherapy,
\newblock Acta Oncologica {\bf 47}, 1286--1293 (2008).

\bibitem{yang2008fast}
D.~Yang, H.~Li, D.~A. Low, J.~O. Deasy, and I.~El~Naqa,
\newblock A fast inverse consistent deformable image registration method based on symmetric optical flow computation,
\newblock Phys. Med. Biol. {\bf 53}, 6143 (2008).

\bibitem{kybic2003fast}
J.~Kybic and M.~Unser,
\newblock Fast parametric elastic image registration,
\newblock TIP {\bf 12}, 1427--1442 (2003).

\bibitem{davatzikos1997spatial}
C.~Davatzikos,
\newblock Spatial transformation and registration of brain images using elastically deformable models,
\newblock Computer Vision and Image Understanding {\bf 66}, 207--222 (1997).

\bibitem{vercauteren2009diffeomorphic}
T.~Vercauteren, X.~Pennec, A.~Perchant, and N.~Ayache,
\newblock Diffeomorphic demons: Efficient non-parametric image registration,
\newblock NeuroImage {\bf 45}, S61--S72 (2009).

\bibitem{vercauteren2007non}
T.~Vercauteren, X.~Pennec, A.~Perchant, and N.~Ayache,
\newblock Non-parametric diffeomorphic image registration with the demons algorithm,
\newblock in {\em MICCAI}, pages 319--326, Springer, 2007.

\bibitem{rueckert2006diffeomorphic}
D.~Rueckert, P.~Aljabar, R.~A. Heckemann, J.~V. Hajnal, and A.~Hammers,
\newblock Diffeomorphic registration using B-splines,
\newblock in {\em MICCAI}, pages 702--709, Springer, 2006.

\bibitem{klein2007evaluation}
S.~Klein, M.~Staring, and J.~P. Pluim,
\newblock Evaluation of optimization methods for nonrigid medical image registration using mutual information and B-splines,
\newblock TIP {\bf 16}, 2879--2890 (2007).

\bibitem{de2019deep}
B.~D. De~Vos, F.~F. Berendsen, M.~A. Viergever, H.~Sokooti, M.~Staring, and I.~I{\v{s}}gum,
\newblock A deep learning framework for unsupervised affine and deformable image registration,
\newblock Med. Image Anal. {\bf 52}, 128--143 (2019).

\bibitem{fu2020deep}
Y.~Fu, Y.~Lei, T.~Wang, W.~J. Curran, T.~Liu, and X.~Yang,
\newblock Deep learning in medical image registration: a review,
\newblock Phys. Med. Biol. {\bf 65}, 20TR01 (2020).

\bibitem{haskins2020deep}
G.~Haskins, U.~Kruger, and P.~Yan,
\newblock Deep learning in medical image registration: a survey,
\newblock Machine Vision and Applications {\bf 31}, 8 (2020).

\bibitem{balakrishnan2019voxelmorph}
G.~Balakrishnan, A.~Zhao, M.~R. Sabuncu, J.~Guttag, and A.~V. Dalca,
\newblock Voxelmorph: a learning framework for deformable medical image registration,
\newblock TIP {\bf 38}, 1788--1800 (2019).

\bibitem{hoffmann2021synthmorph}
M.~Hoffmann, B.~Billot, D.~N. Greve, J.~E. Iglesias, B.~Fischl, and A.~V. Dalca,
\newblock SynthMorph: learning contrast-invariant registration without acquired images,
\newblock TIP {\bf 41}, 543--558 (2021).

\bibitem{chen2022transmorph}
J.~Chen, E.~C. Frey, Y.~He, W.~P. Segars, Y.~Li, and Y.~Du,
\newblock Transmorph: Transformer for unsupervised medical image registration,
\newblock Med. Image Anal. {\bf 82}, 102615 (2022).

\bibitem{zajkac2023ground}
H.~D. Zaj{\k{a}}c, N.~R. Avlona, F.~Kensing, T.~O. Andersen, and I.~Shklovski,
\newblock Ground Truth Or Dare: Factors Affecting The Creation Of Medical Datasets For Training AI,
\newblock in {\em AAAI}, pages 351--362, 2023.

\bibitem{guo2021unsupervised}
Y.~Guo, L.~Bi, D.~Wei, L.~Chen, Z.~Zhu, D.~Feng, R.~Zhang, Q.~Wang, and J.~Kim,
\newblock Unsupervised Landmark Detection-Based Spatiotemporal Motion Estimation for 4-D Dynamic Medical Images,
\newblock IEEE Trans. Cybern {\bf 53}, 3532--3545 (2021).

\bibitem{guo2020spatiotemporal}
Y.~Guo, L.~Bi, E.~Ahn, D.~Feng, Q.~Wang, and J.~Kim,
\newblock A spatiotemporal volumetric interpolation network for 4d dynamic medical image,
\newblock in {\em CVPR}, pages 4726--4735, 2020.

\bibitem{wei2023mpvf}
T.-T. Wei, C.~Kuo, Y.-C. Tseng, and J.-J. Chen,
\newblock MPVF: 4D Medical Image Inpainting by Multi-Pyramid Voxel Flows,
\newblock IEEE J. Biomed. Health Inform.  (2023).

\bibitem{kim2024data}
J.~Kim, H.~Yoon, G.~Park, K.~Kim, and E.~Yang,
\newblock Data-Efficient Unsupervised Interpolation Without Any Intermediate Frame for 4D Medical Images,
\newblock arXiv preprint arXiv:2404.01464  (2024).

\bibitem{molaei2023implicit}
A.~Molaei, A.~Aminimehr, A.~Tavakoli, A.~Kazerouni, B.~Azad, R.~Azad, and D.~Merhof,
\newblock Implicit neural representation in medical imaging: A comparative survey,
\newblock in {\em ICCV}, pages 2381--2391, 2023.

\bibitem{sitzmann2020implicit}
V.~Sitzmann, J.~Martel, A.~Bergman, D.~Lindell, and G.~Wetzstein,
\newblock Implicit neural representations with periodic activation functions,
\newblock NeurIPS {\bf 33}, 7462--7473 (2020).

\bibitem{wolterink2022implicit}
J.~M. Wolterink, J.~C. Zwienenberg, and C.~Brune,
\newblock Implicit neural representations for deformable image registration,
\newblock in {\em MIDL}, pages 1349--1359, PMLR, 2022.

\bibitem{hisham2015template}
M.~Hisham, S.~N. Yaakob, R.~Raof, A.~A. Nazren, and N.~Wafi,
\newblock Template matching using sum of squared difference and normalized cross correlation,
\newblock in {\em 2015 IEEE student conference on research and development (SCOReD)}, pages 100--104, IEEE, 2015.

\bibitem{zhao2006image}
F.~Zhao, Q.~Huang, and W.~Gao,
\newblock Image matching by normalized cross-correlation,
\newblock in {\em 2006 IEEE international conference on acoustics speech and signal processing proceedings}, volume~2, pages II--II, IEEE, 2006.

\bibitem{konig2014fast}
L.~K{\"o}nig and J.~R{\"u}haak,
\newblock A fast and accurate parallel algorithm for non-linear image registration using normalized gradient fields,
\newblock in {\em 2014 IEEE 11th international symposium on biomedical imaging (ISBI)}, pages 580--583, IEEE, 2014.

\bibitem{vishnevskiy2016isotropic}
V.~Vishnevskiy, T.~Gass, G.~Szekely, C.~Tanner, and O.~Goksel,
\newblock Isotropic total variation regularization of displacements in parametric image registration,
\newblock IEEE transactions on medical imaging {\bf 36}, 385--395 (2016).

\bibitem{zhang1994iterative}
Z.~Zhang,
\newblock Iterative point matching for registration of free-form curves and surfaces,
\newblock IJCV {\bf 13}, 119--152 (1994).

\bibitem{niklaus2020softmax}
S.~Niklaus and F.~Liu,
\newblock Softmax splatting for video frame interpolation,
\newblock in {\em CVPR}, pages 5437--5446, 2020.

\bibitem{ode1}
J.~Xu, E.~Z. Chen, X.~Chen, T.~Chen, and S.~Sun,
\newblock Multi-scale neural odes for 3d medical image registration,
\newblock in {\em MICCAI}, pages 213--223, Springer, 2021.

\bibitem{ode2}
Y.~Wu, T.~Z. Jiahao, J.~Wang, P.~A. Yushkevich, M.~A. Hsieh, and J.~C. Gee,
\newblock Nodeo: A neural ordinary differential equation based optimization framework for deformable image registration,
\newblock in {\em CVPR}, pages 20804--20813, 2022.

\bibitem{ode3}
X.~Liang, S.~Lin, F.~Liu, D.~Schreiber, and M.~Yip,
\newblock ORRN: An ODE-based Recursive Registration Network for Deformable Respiratory Motion Estimation With Lung 4DCT Images,
\newblock TBME  (2023).

\bibitem{lddmm}
M.~F. Beg, M.~I. Miller, A.~Trouv{\'e}, and L.~Younes,
\newblock Computing large deformation metric mappings via geodesic flows of diffeomorphisms,
\newblock IJCV {\bf 61}, 139--157 (2005).

\bibitem{hildebrand1987introduction}
F.~B. Hildebrand,
\newblock {\em Introduction to numerical analysis},
\newblock Courier Corporation, 1987.

\bibitem{castillo2009framework}
R.~Castillo, E.~Castillo, R.~Guerra, V.~E. Johnson, T.~McPhail, A.~K. Garg, and T.~Guerrero,
\newblock A framework for evaluation of deformable image registration spatial accuracy using large landmark point sets,
\newblock Physics in Medicine \& Biology {\bf 54}, 1849 (2009).

\bibitem{wasserthal2023totalsegmentator}
J.~Wasserthal et~al.,
\newblock TotalSegmentator: robust segmentation of 104 anatomic structures in CT images,
\newblock Radiology: Artificial Intelligence {\bf 5} (2023).

\bibitem{hodson2022root}
T.~O. Hodson,
\newblock Root mean square error (RMSE) or mean absolute error (MAE): When to use them or not,
\newblock Geoscientific Model Development Discussions {\bf 2022}, 1--10 (2022).

\bibitem{johnson2006signal}
D.~H. Johnson,
\newblock Signal-to-noise ratio,
\newblock Scholarpedia {\bf 1}, 2088 (2006).

\bibitem{brunet2011mathematical}
D.~Brunet, E.~R. Vrscay, and Z.~Wang,
\newblock On the mathematical properties of the structural similarity index,
\newblock IEEE Transactions on Image Processing {\bf 21}, 1488--1499 (2011).

\bibitem{datteri2012estimation}
R.~D. Datteri and B.~M. Dawant,
\newblock Estimation and reduction of target registration error,
\newblock in {\em Medical Image Computing and Computer-Assisted Intervention--MICCAI 2012: 15th International Conference, Nice, France, October 1-5, 2012, Proceedings, Part III 15}, pages 139--146, Springer, 2012.

\bibitem{sharp2010plastimatch}
G.~C. Sharp, R.~Li, J.~Wolfgang, G.~Chen, M.~Peroni, M.~F. Spadea, S.~Mori, J.~Zhang, J.~Shackleford, and N.~Kandasamy,
\newblock Plastimatch: an open source software suite for radiotherapy image processing,
\newblock in {\em Proceedings of the XVI’th International Conference on the use of Computers in Radiotherapy (ICCR)}, volume~3, 2010.

\bibitem{avants2009advanced}
B.~B. Avants et~al.,
\newblock Advanced normalization tools (ANTS),
\newblock Insight j {\bf 2}, 1--35 (2009).

\bibitem{liang2023orrn}
X.~Liang, S.~Lin, F.~Liu, D.~Schreiber, and M.~Yip,
\newblock ORRN: An ODE-based recursive registration network for deformable respiratory motion estimation with lung 4DCT images,
\newblock IEEE Transactions on Biomedical Engineering {\bf 70}, 3265--3276 (2023).

\bibitem{wu2022nodeo}
Y.~Wu, T.~Z. Jiahao, J.~Wang, P.~A. Yushkevich, M.~A. Hsieh, and J.~C. Gee,
\newblock Nodeo: A neural ordinary differential equation based optimization framework for deformable image registration,
\newblock in {\em CVPR}, pages 20804--20813, 2022.

\bibitem{4dlung}
G.~D. Hugo, E.~Weiss, W.~C. Sleeman, S.~Balik, P.~J. Keall, J.~Lu, and J.~F. Williamson,
\newblock Data from 4D lung imaging of NSCLC patients,
\newblock The Cancer Imaging Archive {\bf 10}, K9 (2016).

\bibitem{fu2018adaptive}
Y.~Fu, S.~Liu, H.~H. Li, H.~Li, and D.~Yang,
\newblock An adaptive motion regularization technique to support sliding motion in deformable image registration,
\newblock Medical physics {\bf 45}, 735--747 (2018).

\bibitem{murr2023applicability}
M.~Murr, K.~K. Brock, M.~Fusella, N.~Hardcastle, M.~Hussein, M.~G. Jameson, I.~Wahlstedt, J.~Yuen, J.~R. McClelland, and E.~V. Osorio,
\newblock Applicability and usage of dose mapping/accumulation in radiotherapy,
\newblock Radiotherapy and Oncology {\bf 182}, 109527 (2023).

\bibitem{criscuolo2025vessel}
E.~R. Criscuolo, Z.~Zhang, Y.~Hao, and D.~Yang,
\newblock A vessel bifurcation landmark pair dataset for abdominal CT deformable image registration (DIR) validation,
\newblock Medical Physics  (2025).

\end{thebibliography}
